# Non-genetic acoustic stimulation of single neurons by a tapered fiber optoacoustic emitter

*Linli Shi[1#], Ying Jiang[2#], Fernando R. Fernandez[2,5,6], Lu Lan[3], Guo Chen[3], Heng-ye Man[4,5], John A. White[2,5,6], Ji-Xin Cheng[2,3*], Chen Yang[1, 3*]*

[1] *Department of Chemistry, Boston University, 580 Commonwealth Avenue, Boston, MA 02215, USA*

[2] *Department of Biomedical Engineering, Boston University, 44 Cummington Mall, Boston, MA 02215, USA*

[3] *Department of Electrical and Computer Engineering, 8 St. Mary's Street, Boston, MA 02215, USA*

[4] *Department of Biology, Boston University, 5 Cummington Mall, Boston, MA, 02215, USA*

[5] *Center for Systems Neuroscience, Boston University, 610 Commonwealth Ave, Boston, MA 02215, USA*

[6] *Neurophotonics Center, Photonics Center, Boston University, 8 St. Mary's Street, Boston MA 02215, USA# equal contributions.*

***** *Corresponding author: cheyang@bu.edu, jxcheng@bu.edu*

## Abstract

As an emerging technology, transcranial focused ultrasound has been demonstrated to successfully evoke motor responses in mice, rabbits, and sensory/motor responses in humans. Yet, the spatial resolution of ultrasound does not allow for high-precision stimulation. Here, we developed a tapered fiber optoacoustic emitter (TFOE) for optoacoustic stimulation of neurons with an unprecedented spatial resolution of 20 microns, enabling selective activation of single neurons or subcellular structures, such as axons and dendrites. A single acoustic pulse of 1 microsecond converted by the TFOE from a single laser pulse of 3 nanoseconds is shown as the shortest acoustic stimuli so far for successful neuron activation. The highly localized ultrasound generated by the TFOE made it possible to integrate the optoacoustic stimulation and highly stable patch clamp recording on single neurons. Direct measurements of electrical response of single neurons to acoustic stimulation, which is difficult for conventional ultrasound stimulation, have been demonstrated for the first time. By coupling TFOE with ex vivo brain slice electrophysiology, we unveil cell-type-specific response of excitatory and inhibitory neurons to acoustic stimulation. These results demonstrate that TFOE is a non-genetic single-cell and sub-cellular modulation technology, which could shed new insights into the mechanism of neurostimulation.

## Introduction

Neuromodulation at high spatial resolution poses great significance in advancing fundamental knowledge in the field of neuroscience, as firing of a small population or even single neurons can specifically alter animal behavior or brain state[1,2]. Clinically, precise neural stimulation lays the foundation for procedures such as retinal stimulation[3,4] and selective dorsal rhizotomy[5], where selective activation of a small population or single neurons and axon fibers is desired. Historically, electrical stimulation has been the most important technique for neuromodulation. Deep brain stimulation, as the most prescribed neuromodulation method clinically, has been used for treating neurological and psychiatric disorders, such as Parkinson's Disease, depression, and epilepsy[6-8]. However, the spatial resolution of electrical stimulation is limited by the spread of the electric current, which could distribute over several millimeters and outside of the area of targeting[9]. Providing high spatial precision and cell specificity, optogenetics has been shown as a powerful method of modulating population neural activities in rodents [10,11]. Yet, the requirement of viral infection makes it challenging to be applied in humans. Toward non-genetic stimulation, photothermal neural stimulations based on light absorption of water has been reported [12-14], and it has attracted increasing interest in basic science and translational application[15,16]. In infrared photothermal neural stimulation (INS), near-infrared light between 1.5 and 2 μm in wavelength is delivered through a fiber and converted into temperature increase in water with sub-millimeter precision[15,17], where the associated heating raises a significant concern of tissue damage[18]. As a rapidly growing modality, focused ultrasound has been harnessed in a myriad of brain neuromodulation applications[19,20,21], given its non-invasive nature with a deep penetration depth[22]. However, ultrasound, with a focus limited by the acoustic wave diffraction, offers a limited spatial resolution at the level of several millimeters[19], which hinders the study of specific brain regions.

Additionally, since the ultrasound field easily disrupts the gigaOhm seals[23], it is challenging to integrate ultrasound stimulation with whole-cell patch-clamp electrophysiology, which is the gold standard technique for high-fidelity analysis of the biophysical mechanisms of neural membrane and ion channels[24].

Our team recently developed a fiber-based optoacoustic converter, which exploited the optoacoustic effect, absorbing pulsed light and producing an ultrasound wave[25], and achieved neural stimulation in vitro and in vivo at submillimeter spatial resolution[26]. Yet, such resolution is still insufficient for targeting subtypes of neuron at single cell level or sub-cellular structures. In addition, the device does not allow stable integration with patch clamp on the same cell being stimulated. New capabilities, including single and subcellular precision and integration of single cell electrophysiology recording, are still sought to enable understanding of mechanical stimulation at the single cell level and to offer high precision for potential clinical applications.

Here, we report a miniaturized tapered fiber optoacoustic emitter (TFOE) capable of generating >57 kPa pressure with a spatial confinement around 20 μm, which offers an unprecedented high spatial resolution for ultrasound stimulation. The significant advancement of TFOE in both spatial resolution and optoacoustic conversion efficacy are achieved based on the following innovative designs. First, instead of using a commercial multimode fiber with a diameter of 200 microns as in our earlier work, we developed a controlled tapering strategy and reproducibly tapered the fibers to a tip diameter as small as 20 μm. Second, a new deposition method was developed to achieve uniform and controllable coating thickness of ~10 micron on the small 20-micron fiber tip. Third, instead of using graphite powder in epoxy as a converter, we applied carbon nanotubes (CNT) embedded in a Polydimethylsiloxane (PDMS) matrix with improved solubility, which allows highly efficient optoacoustic signal generation from the tapered fiber tip with an increase in the conversion efficiency by one order of magnitude[27] and prevents light leak from the thin 20 micron coating.

Using TFOE, we pushed spatial and temporal resolution of neuron stimulation. Specifically, we demonstrated single cell stimulations and subcellular stimulation of axons and dendrites. We also showed single acoustic pulse with duration of 1 microsecond was capable to achieve neuron stimulation, which was found as the shortest duration of acoustic stimuli to the best of our knowledge[23]. Importantly, the near field acoustic wave generated by TFOE allowed optoacoustic stimulation with simultaneously monitoring cell response using whole cell patch clamp recording, which had been reported as a challenge for traditional ultrasound[23]. Our studies revealed cell-type-specific response to acoustic stimulation for excitatory and inhibitory neurons. These advances show the exciting potential of TFOE as a platform technology for non-genetic high-precision stimulation of the neural system, and as a tool for investigations into the mechanisms of acoustic neural stimulation and MRI compatible clinical application.

## Results

### Fabrication of TFOE and characterization of acoustic generation

Towards single-cell modulation, we have fabricated a TFOE with a 20 μm tip diameter as a miniaturized low-frequency ultrasound source. We took several innovative steps to overcome the challenges associated with the small 20 μm fiber tip. For control of tapering an optical fiber reproducibly, a multimode fiber was gradually pulled from the full diameter of 225 μm to 20 μm via a thermal tapering technique (see **Methods**). To convert the light energy into acoustic waves with maximum efficiency[28], we have optimized the absorption/thermal expansion layer, which composes multi-wall carbon nanotubes (CNTs) with strong light absorption embedded in PDMS with a high thermal expansion coefficient[27]. To increase optoacoustic conversion efficiency in the tapered fiber and assure minimum light leakage, the optoacoustic CNT/PDMS coating was prepared with a large CNT concentration of 15%, by introducing isopropyl alcohol (IPA) to form IPA-coated CNTs with hydroxyl groups. To overcome the reduced viscosity of PDMS induced by high CNT concentration and IPA, as well as to achieve a uniform and controlled coating thickness on the 20 μm cross-section of the tapered end, a punch-through method (see **Methods**) was deployed (**Fig. 1b**). The coating thickness was controlled by changing the matrix viscosity via IPA evaporation. The tapered fiber optoacoustic emitter was further confirmed by microscope imaging to have a CNT/PDMS coating of a thickness of 9.5 μm and an overall diameter of 19.8 μm, meeting the needs of single-cell targeting (**Fig. 1b bottom**).

Next, a 1030 nm nanosecond pulsed laser was delivered to the TFOE to generate optoacoustic signals. **Fig. 1c** shows the average trace of acoustic waveform. The near field ultrasound pressure was 56.7 kPa, measured by a needle hydrophone. Radio frequency spectrum of the acoustic waveform after Fast Fourier Transform (FFT) exhibits a broadband acoustic frequency from 0 to 10 MHz with a peak frequency at 1.0 MHz (**Fig. 1d**), which is within the most frequently used range for transcranial in vivo neuromodulation[29] as well as other biomedical applications[30]. Due to the intrinsic acoustic diffraction limit, the conventional transducer array is not able to characterize the localization of the acoustic field with sufficient spatial resolution. Alternatively, the motions of fluorescent Polymethylmethacrylate (PMMA) beads (Dia. 9 μm) dispersed in Phosphate-buffered saline (PBS) under the application of the TFOE were used to visualize the distribution of acoustic field (**Fig. S1**). Under a laser burst duration of 50 milliseconds and laser power of 7.8 mW with repetition rate of 1.7 kHz, two beads (dashed circles, **Fig. S1**) close to the TFOE tip (less than 1 μm) showed a displacement of 5 μm, while other beads ~20 μm away from the TFOE tip remained stationary upon the TFOE treatment, indicating that the acoustic field generated by the TFOE is localized within a distance of 20 µm.

To characterize the thermal profile generated by TFOE in water during acoustic generation, temperature on the fiber tip was measured by a miniaturized ultrafast thermal sensor (DI-245, DataQ, OH, USA) directly in contact with the TFOE tip surface. Two test conditions were used for successful neuron stimulation: first, a laser pulse train of 50 milliseconds, laser power at 7.8 mW and a repetition rate of 1.7 kHz; second, a laser pulse train of 1 millisecond, a laser power at 11.4 mW and a repetition rate of 1.7 kHz. As shown in **Fig. 1.4e,** the tip surface temperature increased by only 0.093±0.004 °C under the first condition and increase was not detectable under the second condition. This temperature increase is far below the threshold of thermal induced neuron modulation ($\Delta T \geq 5$ °C)[31,32]. Collectively, these results demonstrate that TFOE with a tip diameter of 20 µm fabricated serves as a point ultrasound source, producing ultrasound fields that are highly localized to around 20 µm from the tip. This unprecedented spatial confinement will enable high precision stimulation at single neuron level while minimizing thermal damage and undesired mechanical disruptions.

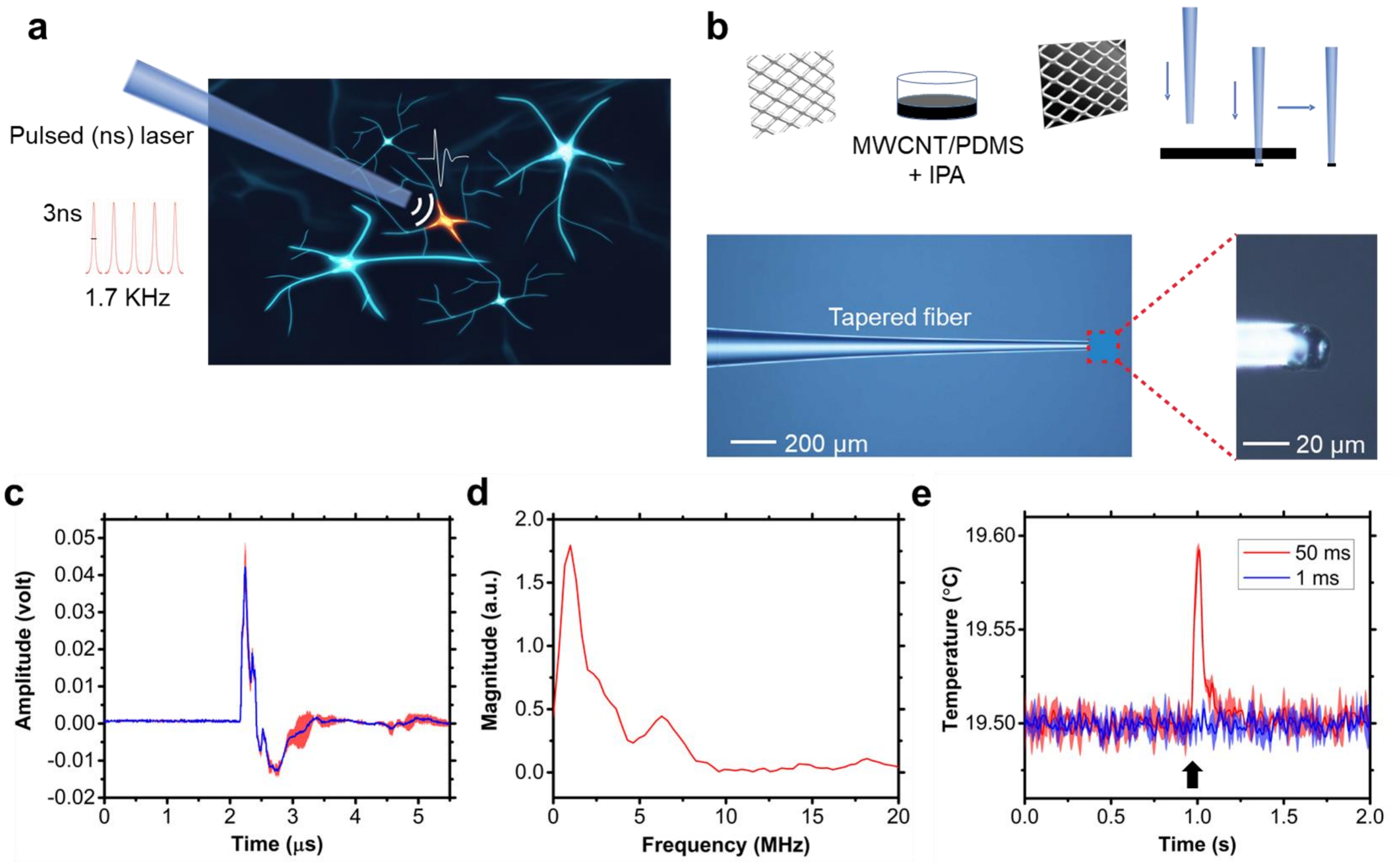

**Figure 1. TFOE for high-precision optoacoustic stimulation. (a)** Schematic of TFOE enabling single neuron stimulation. **(b)** multiwall CNT/PDMS mixture as coating material casted on a metal mesh followed by a punch-through method to coat the tapered fiber. Bottom left: optical images of TFOE, Bottom right: zoom-in showing the tip. **(c, d)** Representative acoustic signal generated by TFOE in time domain (c) and frequency domain (d). Shaded area in (c): standard deviation taken from three measurements from the same TFOE. **(e)** Surface temperature of TFOE tip during laser excitation of 50 milliseconds (7.8 mW, Red) and 1 millisecond (11.4 mW, Blue), respectively. Shaded area in (e): standard deviation taken from three measurements from the same TFOE. Black arrow: laser onset.

## TFOE stimulation of primary neurons with single cell precision

To test whether the TFOE provides sufficient spatial precision when modulating a single neuron in culture, we prepared primary rat cortical neurons expressing GCaMP6f and performed calcium imaging using an inverted wide field fluorescence microscope. Controlled by a micro-manipulator, a TFOE was placed approximately 5 μm away from a targeted neuron. A 3-nanosecond pulsed laser at 1030 nm and 1.7 kHz repetition rate was used to deliver laser pulses of 50 milliseconds duration at an average power of 7.8 mW, corresponding to 85 pulses. Calcium transients were observed immediately after laser onset for the targeted neurons, while other neurons approximately 50 μm to 70 μm away from the tip remained unaffected (**Fig. 2a**), indicating high spatial resolution of TFOE stimulation. The calcium transient with max ΔF/F of 135%±83% (N = 6 from 3 cultures, data in mean ± SD) indicates successful activation of the targeted neuron likely through firing of multiple action potentials evoked by TFOE stimulation. To further improve the temporal resolution, a laser pulse train of 1 millisecond (2 pulses) at 11.4 mW power was delivered to the TFOE. Successful activation of single neurons was also observed with a max ΔF/F of 106%±61% (N = 8 cells from 3 cultures, data in mean ± SD) (**Fig. 2b**).

**Figure S2** compared the TFOE stimulation with controls. The control group of 1 millisecond TFOE with 3 μM tetrodotoxin (TTX) showed no activation, confirming that the calcium increase observed in the experimental groups resulted from $Na^+$ channel-dependent action potentials. A laser only group with a pulse train of 1.0 s and 11.4 mW power using a tapered fiber without the coating showed no activation, therefore, the effect of the laser on the neuron activity can be excluded.

We next investigated whether the TFOE can trigger neural activation reliably and repeatedly. **Figure 2c** shows the fluorescence intensity of the same neuron upon repeated TFOE stimulation for three times. We used 1 millisecond laser duration for each stimulation and an interval of 1 min between each recording period. Successful activation was achieved for each stimulation on the same neuron, which confirmed the neuron's viability after TFOE stimulation. A decrease in max ΔF/F for each sequential stimulation was observed, which could be attributed to calcium depletion[33,34] or spike frequency adaptation[35]. In addition, we demonstrated the spatial precision of the TFOE stimulation using three neurons selectively targeted by the TFOE. These three neurons had an edge-to-edge spacing of 25±2 μm. The TFOE was sequentially placed about 5 μm away from each of the three targeted neurons. The maximum fluorescence intensity change (**Δ**F/F) was color-labeled for each neuron in red, yellow and green, respectively (**Fig. 2d**). Importantly, fluorescence increase was observed only for the selectively targeted neuron without simultaneous activation of the other two neurons, indicating that TFOE stimulation provided a spatial resolution of less than 25 μm. These results collectively confirm that TFOE can stimulate single neurons reliably and repeatability.

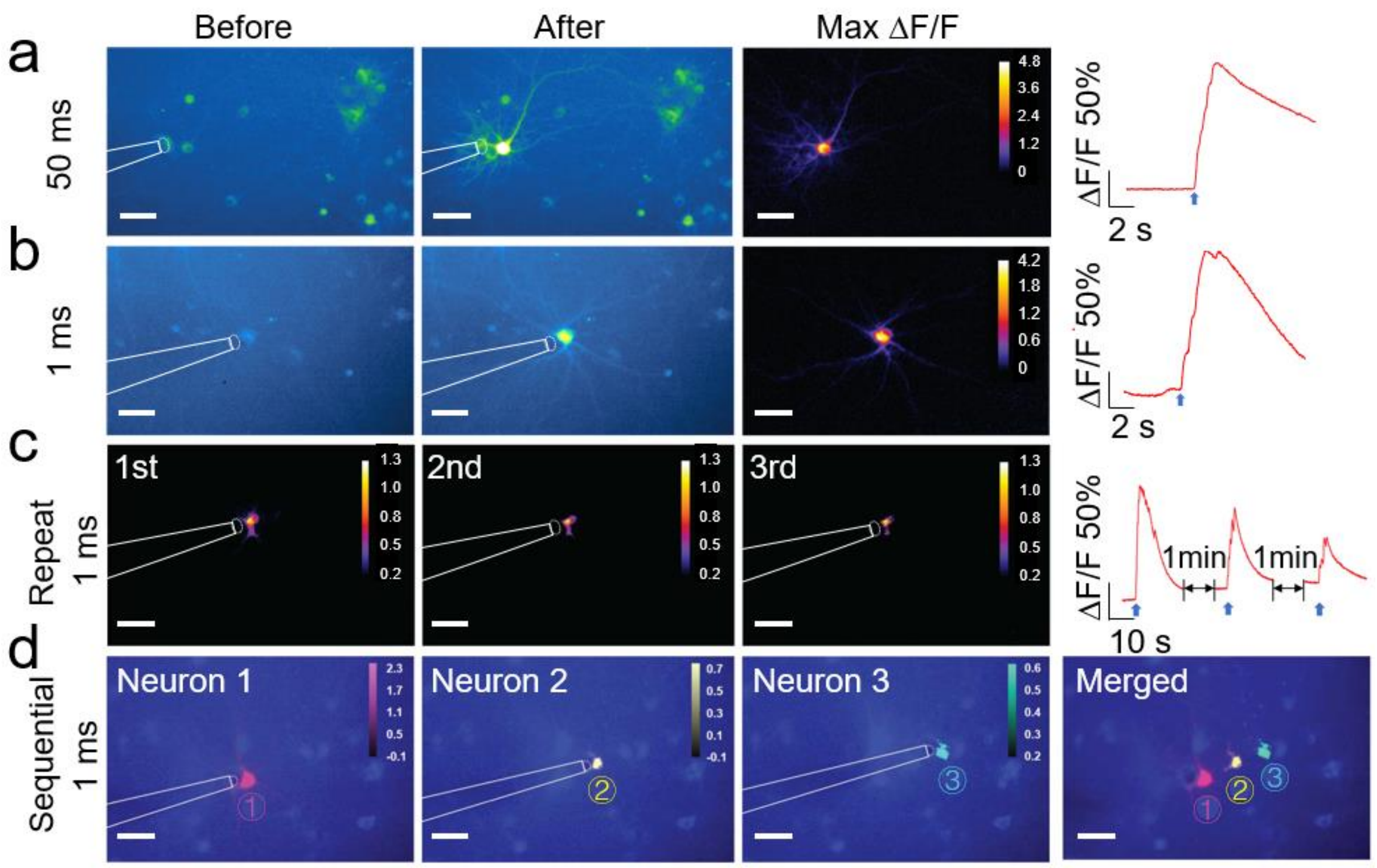

**Figure 2. Fluorescence images of GCaMP6f-expressing neurons in response to TFOE stimulation.** (**a, b**) Fluorescence images and calcium traces in sparse population stimulated by TFOE with a laser duration of 50 milliseconds (a) and 1 millisecond (b). The fluorescence images with peak intensity after stimulation are shown as "after". Blue arrows: laser onset. (**c**) Max ΔF/F images of one neuron undergone repeated TFOE stimulation 3 times. Right: calcium signal of the neuron. Blue arrows: laser onset. (**d**) Sequential stimulation of 3 neurons. Max ΔF/F labeled in red, yellow, green respectively superimposed with the fluorescence image. Scale bars: 50 μm.

**Optoacoustic stimulation with a single pulse**

Taking advantage of the controllability of laser pulse energy and pulse number, we explored the stimulation effect of single optoacoustic pulse on neurons. The same nanosecond pulsed laser was used to deliver a single laser pulse to the TFOE. TFOE stimulation of the GCaMP6f-expressing primary cortical neuron with different laser pulse energy was performed under the single pulse condition. No calcium transient was observed until the pulse energy reached 6 µJ/pulse (**Fig. 3a-d**). The width of the optoacoustic wave is approximately 1 microsecond (**Fig.1c**), which is, to the best of our knowledge, the shortest acoustic stimuli for successful neuron modulation so far[23]. This capability could potentially enable acoustic control of neural circuits with unprecedented temporal precision required to mimic natural neural coding[36].

We further investigated the required laser energy for a given pulse number for successful neuron stimulation. In previous ultrasound studies, continuous wave and pulsed ultrasound with varied intensities and durations has been applied for neuron stimulation[21]. The relationship between temporal-averaged US intensity and response amplitude or success rate was found to be negative[37] or positive[38]. Given these studies, we pursued a statistical investigation of the behavior of neurons in response to acoustic stimulation across multiple intensities and durations. In our work, first, the threshold of pulse energy for successful stimulation is defined as the laser pulse energy sufficient to induce a maximum fluorescence intensity change (Max **Δ**F/F) greater than 20%, since Max **Δ**F/F of 20% GCaMP6f was identified to correspond to ≥ 1 action potential[39]. The threshold energy shows a monotonic decrease from 6.3 µJ, 4.9 µJ to 3.9 µJ when increasing the pulse number from 1, 2 to 4, respectively, and it remains relatively constant at 3.9 µJ and 3.6 µJ when the pulse number increased to 6 and 8, respectively. These results demonstrated the following key findings. First, the decrease of the energy threshold when the laser pulse number increases in the range of 1 to 4 shows that under the small pulse energy condition, subthreshold depolarizations accumulate with increasing pulse numbers, consistent with previous work[40]. Second, the flattening trend of the threshold energy from 4 to 8 pulses implies a presence of an energy threshold at around 4 µJ/pulse, below which the action potential can hardly be evoked with even further elongation of the pulse train. These results are in agreement with previous work[40,41].

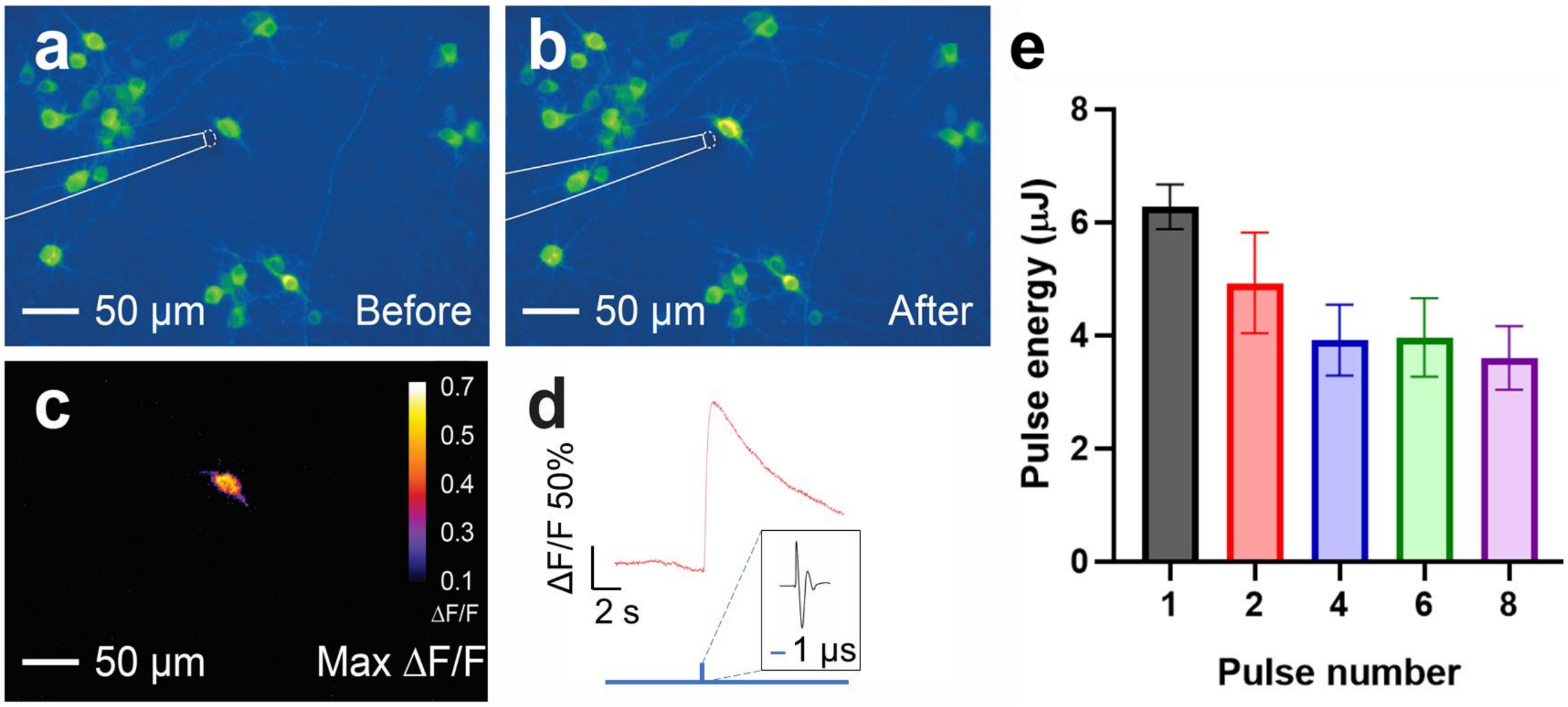

**Figure 3. Pulse energy dependence of TFOE stimulation. (a-c)** Fluorescence images of GCaMP6f expressing neurons before and after TFOE stimulation with a single pulse. **(d)** Calcium trace of the targeted neuron undergone single pulse stimulation. Blue line: onset of optoacoustic stimulation with zoom-in showing a representative optoacoustic waveform. **(e)** Pulse energy threshold for successful neuron stimulation as a function of pulse number (N=5-7).

**TFOE stimulation targeting sub-cellular region of a single neuron**

Upon successful stimulation of cultured primary neurons, we further investigated whether the TFOE can target subcellular structures. To this end, the TFOE was first carefully placed above the targeted area where axons and dendrites densely populate without the presence of somas. A 1030-nm laser pulse train with a duration of 1 millisecond, a laser power of 11.4 mW and a repetition rate of 1.7 kHz was delivered to the TFOE. An increase in fluorescence intensity at the targeted area was clearly observed after laser onset, indicating successful TFOE stimulation of targeted neurites (**Fig. 4a-b**). Three different calcium dynamics were captured through imaging throughout the field of view. First, a slow propagation of calcium wave initiating from the targeted region in the neural network was observed after TFOE stimulation (**Fig. S3**). The speed of the calcium wave propagation was calculated to be 75.2 μm/s, which was in agreement with the propagation speed of dendritic calcium wave induced by synaptic activity or by activity of metabotropic glutamate receptors (mGluRs) and backpropagating action potentials, which generated a speed of ~70 μm/s[42]. Second, 4 sites in the field of view showed elevated fluorescence signals prior to the spreading of calcium wave (**Fig. 4c**). The neurites in the targeted region (labeled purple in **Fig. 4a-c**) and a specific neuron 1 (labeled cyan in **Fig. 4a-c**) with axon directly connecting to neurites in the targeted region showed fast calcium transients immediately after laser onset (**Fig. 4d**), which resembles backpropagation of action potentials. Considering that an unmyelinated axon would conduct action potential spikes at a speed of 500 μm/milliseconds to synapses[43], the propagation from neurites to neuron 1 (cyan in **Fig. 4a-c**) over a distance of approximately 100 μm only requires 0.2 milliseconds. Therefore, the difference in the calcium transient onset for neuron 1 (cyan in **Fig. 4d**) and the targeted area (purple in **Fig. 4d**) was non-detectable by the camera with a sampling interval 50 milliseconds. Third, neuron 2 and 3 (labeled red in **Fig. 4a-c**) in the vicinity but without axons connecting to the targeted area showed an activation delay of ~0.2 s (**Fig. 4d, inset**) with similar temporal dynamics. This signaling was likely attributed to action potential evoked through synaptic transmission, since it showed a faster propagation speed than the calcium wave.

This capability of TFOE induced stimulation on subcellular structures, specifically on axons and dendrites, was then utilized to elucidate whether axons and dendrite have distinct response profiles to optoacoustic stimulation. In **Figure 4e**, three neurites in a multipolar neuron were targeted selectively by the TFOE.  Targeted TFOE stimulation on one of the neurites (red in **Fig. 4e** and **f**) induced strong calcium activation at the soma with no delay (**Fig. 4f, j**). Thus, this neurite is identified as an axon, since such propagating activation resembles backpropagation of action potentials in an axon. Distinctively, targeted TFOE stimulation of the other two neurites (yellow and blue in **Fig. 4e**, **g** and **h**) did not induce any activation at the soma (**Fig. 4g-j**). Thus, they were identified as dendrites. Neuronal dendrites are known to integrate synaptic inputs from upstream neurons, which involves summation of stimuli that arrive in rapid succession, entailing the aggregation of inputs from separate branches. In our case, the forward propagation of a single dendrite was found to be insufficient to evoke action potentials at the soma. The differences between responses of the axon and dendrites upon acoustic stimulation at the single cell level are shown to be repeatable across multiple neurons (**Fig. S4**). Collectively, these data reveal differential response dynamics of axons and dendrites to optoacoustic stimulation for the first time, enabled by subcellular targeting capability of TFOE.

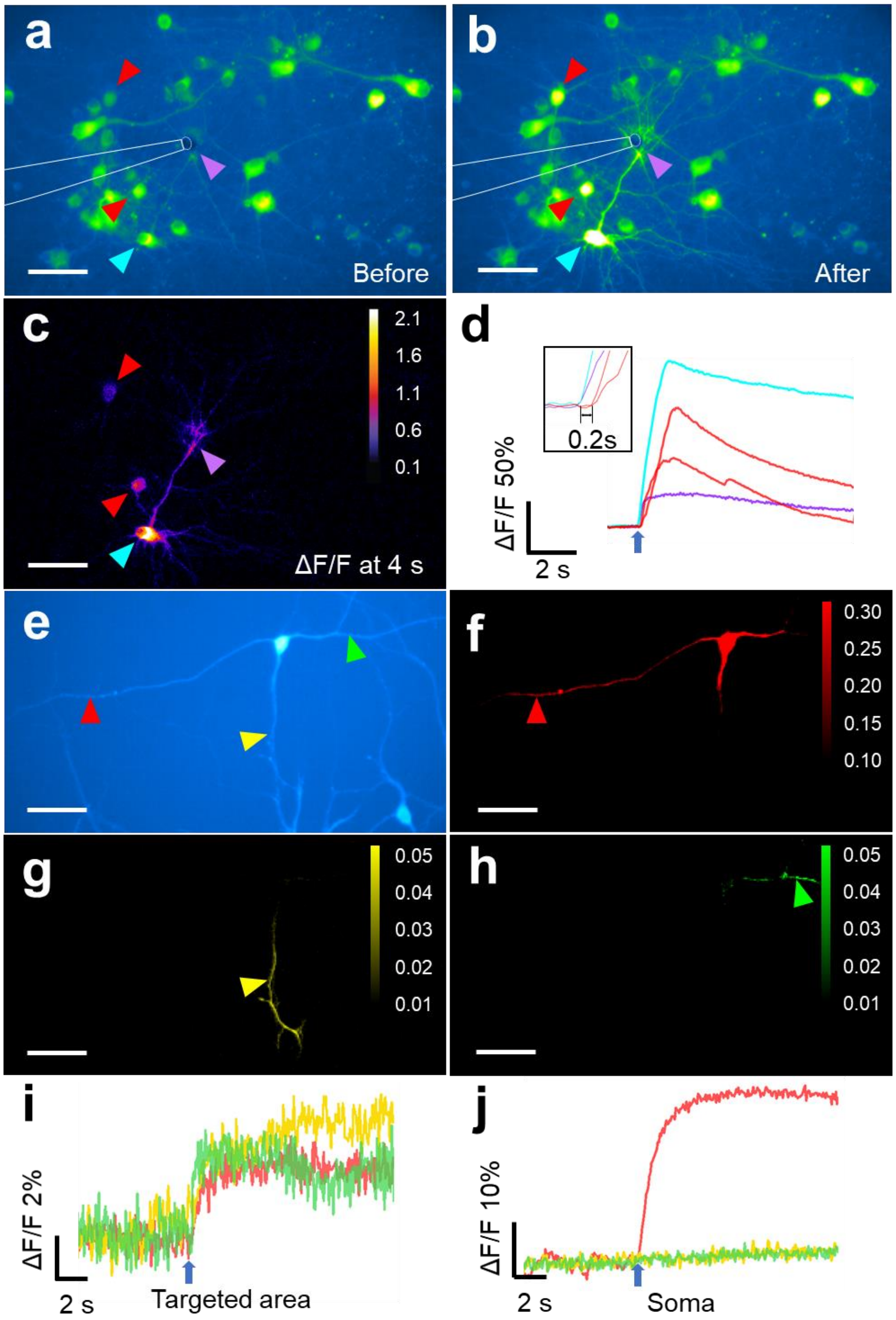

**Figure 4. TFOE evoked sub-cellular stimulation on neurites. (a-b)** TFOE evoked neurites activation with calcium wave propagating along neuron network. Colored arrows: targeted area (purple), neuron 1 (cyan) and neuron 2, 3 (red). **(c)** ΔF/F of calcium signal at 4 s after laser onset. Scale bars: 50 μm. **(d)** Calcium traces of targeted area (purple), neuron 1 (cyan) and neuron 2 and 3 (red), as labeled in (a). Inset: Zoom-in of calcium signals immediately after the laser onset. Black arrow: laser onset. **(e)** A multipolar neuron stimulated with a TFOE selectively targeting the axon

(red arrow) and dendrites (yellow and green arrows). **(f-h)** Maximum ΔF of calcium signal upon stimulation of different areas. **(i)** Calcium traces measured at the targeted neurites as labeled in (e) by red, yellow and green arrows, respectively. **(j)** Calcium traces measured at the soma upon stimulation of different neurites. Blue arrows in (i-j): laser onset.

**Whole cell patch clamp recording reveals cell type specific response to TFOE stimulation**

A key advantage of single-neuron TFOE stimulation is the compatibility with intracellular patch clamp recordings. While the calcium response to the stimulation has limited temporal resolution, direct recordings using intracellular patch clamp recordings stand as the golden standard to study sub- and supra-threshold neuron activity. Conventional ultrasound easily disrupts the patch attachment between the glass and membranes, so intracellular patch clamp recordings have been challenging during conventional ultrasound stimulation. Our optoacoustic stimulation has the advantage of the localized optoacoustic field with a minimized mechanical disruption; therefore, it can be recorded with patching, providing a new testing system to gain insights towards mechanical modulation of neural systems.

We integrated TFOE stimulation with patch clamp recording on single neurons in mouse cortical slices to detect the direct electrical response to optoacoustic single neuron stimulation. As shown in **Figure 5a**, we used brain slices from mice expressing tdTomato in GAD2 interneurons to assist in visualization of specific cell types. Thus, GAD2-tdTomato positive inhibitory interneurons and GAD2-tdTomato negative pyramidal neurons can be selectively targeted. TFOE can be integrated with the patch pipette to induce depolarization leading to action potential generation in the targeted neurons. Also indicated in **Figure 5c-f**, the neuron membrane voltage can be measured precisely with an unprecedented stability upon TFOE stimulation.

For excitatory pyramidal cells, under the current clamp mode, a train of action potential was observed immediately after TFOE stimulation at 5 µm (**Fig. 5c**). The result was consistent with previous calcium imaging with ΔF/F greater than 100% in fluorescence change (**Fig. 2a-b** and **Fig. 3d**). When the TFOE was moved from 5 to 10 µm away from the neurons (**Fig. 5c, d**), the action potentials give way to a subthreshold depolarization, indicating a high confinement of the acoustic field in the tissue.

Next, we targeted tdTomato positive interneurons. TFOE induced subthreshold depolarization in inhibitory interneurons held at -75 mV, and the electrical response over time after stimulation showed two components (**Fig. 5e, inset**). The first sharp peak could be due to the direct interruption of the membrane integrity by the acoustic wave, and the following broad peak is likely due to an inward channel current, thus indicates the possible involvement of ion channels. With the membrane depolarized via injecting positive currents to near -40 mV, a short train of three action potentials was observed upon TFOE stimulation (**Fig. 5f**). The distinct response of excitatory pyramidal neuron and inhibitory interneurons to acoustic stimulation is likely contributed by multiple factors including a unique intrinsic action potential threshold of these two cell types, as well as distribution of mechanosensitive ion channels that have different response dynamics to acoustic radiation force[44,45]. In summary, the TFOE provides an unprecedented stable ultrasound source compatible with patch clamp recordings, holding promise to shed light on the mechanism of acoustic induced neuron stimulation.

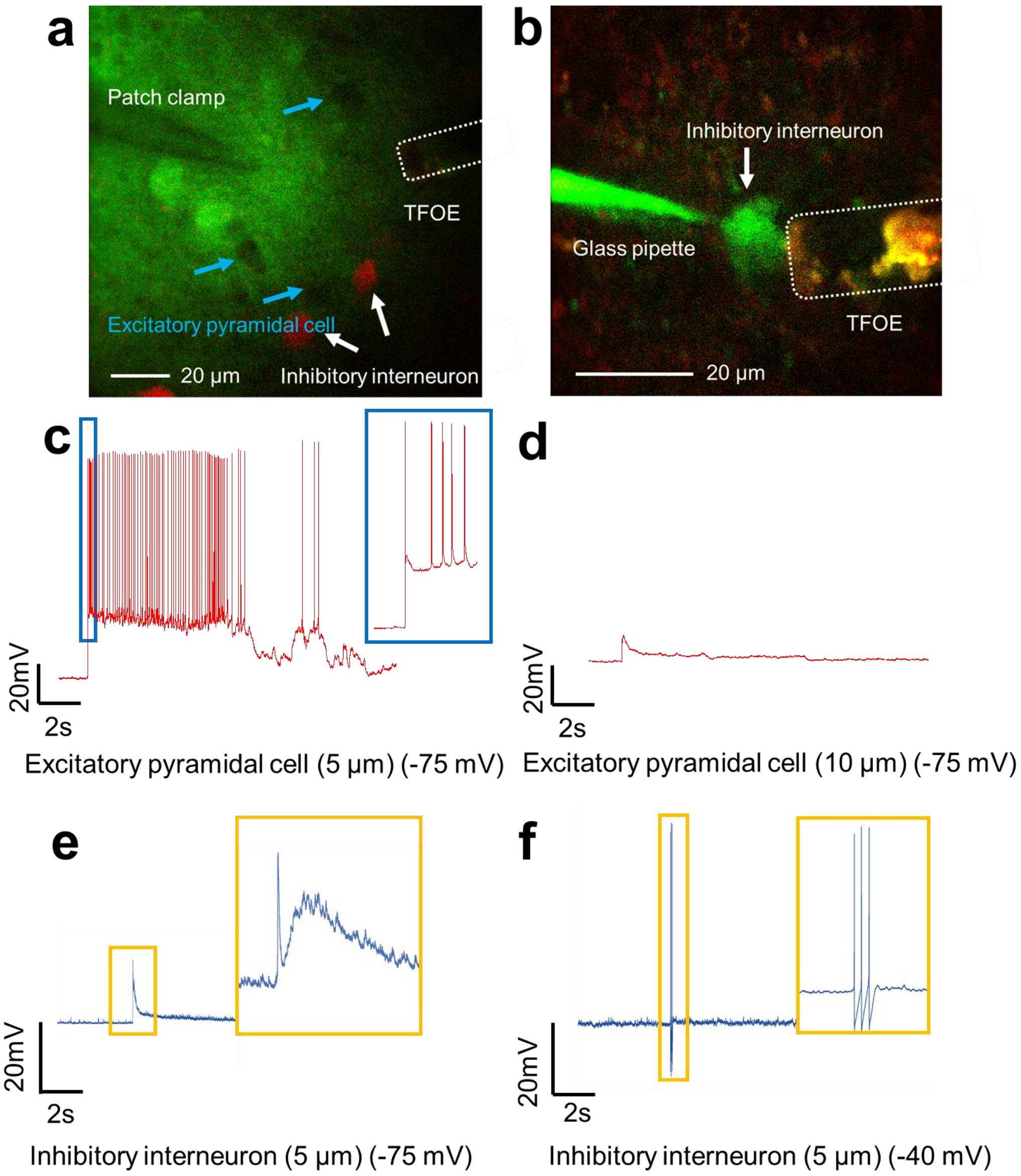

**Figure 5. Single neuron patch clamp with TFOE stimulation. (a-b)** Two photon imaging of patch clamp integrated with TFOE in a mouse brain slice targeting GAD2-tdTomato negative pyramidal neurons and GAD2-tdTomato positive inhibitory interneurons. The patch pipette is visualized using the cyan-green fluorescent dye Alexa Fluor 488 in the intracellular electrode solution **(c, d)** Membrane voltage response in an excitatory pyramidal cell upon TFOE stimulation (5 milliseconds) at a distance of ~5 µm (c) and ~10 µm (d). **(e, f)** Voltage response in an inhibitory interneuron upon TFOE stimulation at ~5 µm at the membrane voltages of -75 mV (e) and -40 mV (f). Laser: 11.4 mW, 1.7 kHz, 5 milliseconds duration.

## Discussion

In this study, we develop a TFOE that generates acoustic waves with a spatial confinement of ~ 20 µm, enabling optoacoustic neural modulation with single neuron and subcellular precision. The near field acoustic wave generated by TFOE allows optoacoustic stimulation along with simultaneous monitoring of cell responses using whole cell patch clamp recording, which has been reported to be challenging under conventional ultrasound stimulation. Coupling TFOE with ex vivo brain slice electrophysiology, we revealed cell-type-specific responses to acoustic stimulation for excitatory and inhibitory neurons.

The optoacoustic effect has been extensively used for biomedical imaging[25], and more recently, it has been explored for neuromodulation [26]. Compared to previously reported optoacoustic stimulator, the TFOE offers new capabilities through adapting new device designs and innovative fabrication methods. The highly efficient optoacoustic conversion layer in the TFOE is made of carbon nanotubes of improved solubility embedded in a thermo-expansive PDMS matrix, which significantly improves light to sound conversion efficiency[27]. Additionally, the punch-through coating method ensures uniform coating of the much smaller tapered fiber tip with great control and reliability.

A key advantage of the TFOE is its unprecedented spatial resolution. Transcranial ultrasound neuromodulation has been demonstrated in rodents[20], non-human primates[46] and in humans[19]. However, due to the wave diffraction limit, focused ultrasound neuromodulation offers a spatial precision of a few millimeters[19], which prohibits site-specific modulation in small animals or single neuron stimulation and therefore lacks the capabilities to study cell-type-specific responses. To overcome this limitation, TFOE generates localized acoustic field at 1 MHz with a spatial resolution of ~ 20 µm, 1000 times smaller compared to the acoustic wavelength. Utilizing the localized acoustic field, we demonstrate neural stimulation with single cell and subcellular precision, and reveal the differential response to TFOE stimulation of subcellular structures by specifically targeting the neuronal soma, dendrites and axons.

By harnessing the controllability of the pulsed laser, we identified the accumulative effect of optoacoustic stimulation at the single cell level, indicating that ultrasonic stimulus can be integrated over a finite duration to become effective. A previous study by Tyler et al. using ultrasound with focal size of 2 mm, the relationship between the temporal-averaged US intensity and the success rate was found to be negative[37]. This differs from Mourad et al.'s study with ultrasound focal size of 1 mm, where positive relationship was reported[38]. These studies provide conflicting evidence[21], where the observed behavior may due to changes in ultrasound parameters, or selective modulation of specific region. In another study by Pauly et al., using an ultrasound with focal spot of 4 mm in diameter was used to induce short-latency muscle contractions in mice measured by electromyography (EMG), showing the ultrasonic stimulus could be integrated over time, and the presence of energy threshold was demonstrated, below which the response could not be evoked with even further elongation of the stimulation duration[40]. The result by Pauly et al. is consistent with the TFOE data. In addition to intrinsic cell properties, their observations could originate from non-specific neural network targeting and recording. The result from TFOE with the capability of assessing single neuron activity in a network-free condition further ascertains the stimulus accumulation effect as an intrinsic signal interpretation of individual neurons.

More importantly, successful TFOE stimulation has been achieved with a single laser pulse of 3 nanoseconds, which generates an acoustic pulse of 1 microsecond. Previously, single tone burst ultrasound with 10 acoustic cycles and overall duration of 22.7 microseconds has been reported as the shortest acoustic stimuli for neuron modulation[23]. Therefore, our result represents significant improvement of temporal resolution of current acoustic stimulation techniques.

Furthermore, TFOE allows integration of acoustic stimulation with whole cell patch clamp recordings. Our electrophysiological recordings of TFOE stimulated single neurons in brain slices revealed distinct responses of excitatory pyramidal neurons and inhibitory interneurons to TFOE stimulation. The distinct responses may originate from differences in the intrinsic threshold or variations in the distribution of different ion channels. Moreover, the inhibitory neurons showed elevated threshold of action potential generation compared to excitatory neurons. This contrasts with finding using electric stimulation, where the inhibitory neurons have a lower threshold than pyramidal cells[47,48]. This discrepancy can be attributed to the different mechanism between acoustic and electric stimulation. Thus, further study, for example, on the ion channel involvement during acoustic stimulation by pharmacologically or genetically modifying ion channels, will provide new insight to the electrophysiological mechanisms of mechanical neuromodulation.

In summary, this genetic-free, single-cell stimulation technique offers a new tool to understand the mechanism of neuron stimulation and how individual neurons work together in networks to implement neural computation. Without any metal components, the TFOE is immune to electromagnetic interference and is compatible with magnetic resonance imaging (MRI)[28], which hold promise for future study toward understanding of behavior and disease in human patients.

## Methods

**Optical fiber tapering.** To control the tapering, a multimode fiber (FT200EMT, Thorlabs, Inc., NJ, USA) was pulled at one end by a traction weight with the other end fixed. The pulling force, determined by the weight of the traction object, was found to be proportional to the square of the tapered end radius, therefore used as the key parameter to control the diameter of the tapered end. In this way, with a pulling force of 0.75 N, tapered fibers of 18.4±0.9 μm (N=5) in diameter were fabricated with high reproducibility.

**Tapered fiber coating.** To assure a maximum optoacoustic conversion efficiency and minimum light leakage in the tapered fiber, CNT/PDMS/IPA composite was prepared. For PDMS, the silicone elastomer (*Sylgard* 184, Dow Corning Corporation, USA) was dispensed directly into a container carefully to minimize air entrapment, followed by mixing with the curing agent in a ratio of 10:1 by weight. Multiwall CNTs (<8 nm OD, 2-5 nm ID, Length 0.5-2 μm, VWR, Inc., NY, USA) and isopropyl alcohol (IPA) were added to PDMS. The mixture was sonicated for 5 min followed by degassing in vacuum for 30 min. Considering the evaporation of IPA, the final CNT concentration in PDMS reached to 15%. The coating matrix was then casted on a metal mesh to form a uniform film. After partial evaporation of the IPA at room temperature for 10 min, the fiber was controlled by a 3-D micromanipulator to punch through the film with a layer transferred to the tapered end. The coated fiber was then cured vertically at 100 °C.

**Optoacoustic signal measurement.** A customized and compact passively Q-switched diode-pumped solid-state laser (1030 nm, 3 ns, 100 μJ, repetition rate of 1.7 kHz, RPMC, Fallon, MO, USA) was used as the excitation source. The laser was connected to an optical fiber through a homemade fiber jumper (SMA-to-SC/PC, ~81% coupling efficiency), then connected to the TFOE with a SubMiniature version A (SMA) connector. To adjust the laser power, fiber optic attenuator sets (multimode, varied gap of 2/4/8/14/26/50 mm, SMA Connector, Thorlabs, Inc., NJ, USA) were used. One miniaturized ultrasound transducer with a central frequency of 5 MHz (XMS-310-B, Olympus, MA, USA) was used to record the optoacoustic signals. The ultrasonic signal was first amplified by an ultrasonic pre-amplifier (0.2–40 MHz, 40 dB gain, Model 5678, Olympus, MA, USA) and then sent to an oscilloscope (DSO6014A, Agilent Technologies, CA, USA) to readout. The signal was measured at 3.5 mm away from the TFOE tip and averaged 100 times with the oscilloscope built in function. The pressure generated was calibrated using a hydrophone with a diameter of 40 µm and frequency range of 1-30 MHz (Precision Acoustics Inc., Dorchester, UK) placed at 1 mm away from the TFOE. All of the instruments were synchronized by the output from the active monitoring photodiode inside the laser. Data analysis and FFT were performed using Origin.

**Embryonic neuron culture.** All experimental procedures have complied with all relevant guidelines and ethical regulations for animal testing and research established and approved by the Institutional animal care and use committee of Boston University. Primary cortical neuron cultures were derived from Sprague-Dawley rats. Cortices were dissected out from embryonic day 18 (E18) rats of either sex and digested in papain (0.5 mg/mL in Earle's balanced salt solution) (Thermo Fisher Scientific Inc., MA). Dissociated cells were washed with and triturated in 10% heat-inactivated fetal bovine serum (FBS, Atlanta Biologicals, GA, USA), 5% heat-inactivated horse serum (HS, Atlanta Biologicals, GA, USA), 2 mM Glutamine-Dulbecco's Modified Eagle Medium (DMEM, Thermo Fisher Scientific Inc., MA, USA), and cultured in cell culture dishes (100 mm diameter) for 30 min at 37 °C to eliminate glial cells and fibroblasts. The supernatant containing neurons was collected and seeded on poly-D-lysine coated cover glass and incubated in a humidified atmosphere containing 5% $CO_2$ at 37 °C with 10% FBS + 5% HS + 2 mM glutamine DMEM. After 16 h, the medium was replaced with Neurobasal medium (Thermo Fisher Scientific Inc., MA, USA) containing 2% B27 (Thermo Fisher Scientific Inc., MA, USA), 1% N2 (Thermo Fisher Scientific Inc., MA, USA), and 2 mM glutamine (Thermo Fisher Scientific Inc.,

MA, USA). Cultures were treated with 5μM FDU (5-Fluoro-2'-deoxyuridine, Sigma-Aldrich, MO, USA) at day 5 in culture to further reduce the number of glial cells. AAV9.Syn.Flex.GCaMP6f.WPRE.SV40 virus (Addgene, MA, USA) was added to the cultures at a final concentration of 1 μl/ml at day 5 in culture for GCaMP6f expressing. Half of the medium was replaced with fresh culture medium every 3-4 days. Cells cultured in vitro for 10-13 days were used for TFOE stimulation experiment.

**In vitro neurostimulation.** A Q-switched 1030-nm nanosecond laser (Bright Solution, Inc. Calgary Alberta, CA) was used to deliver light to the TFOE. A 3-D micromanipulator (Thorlabs, Inc., NJ, USA) was used to position the TFOE approaching the cells. Calcium fluorescence imaging was performed on a lab-built wide-field fluorescence microscope based on an Olympus IX71 microscope frame with a 20x air objective (UPLSAPO20X, 0.75NA, Olympus, MA, USA), illuminated by a 470 nm LED (M470L2, Thorlabs, Inc., NJ, USA) and a dichroic mirror (DMLP505R, Thorlabs, Inc., NJ, USA). Image sequences were acquired with a scientific CMOS camera (Zyla 5.5, Andor) at 20 frames per second. Neurons expressing GCaMP6f at DIV (day in vitro) 10-13 were used for stimulation experiment. For TTX control group, tetrodotoxin citrate (ab120055, Abcam, MA, USA) was added to the culture to reach 3 μM final concentration 10 min before Calcium imaging. The fluorescence intensities, data analysis and exponential curve fitting were analyzed using ImageJ (Fiji) and Origin 2019.

**Ex-vivo whole cell patch clamp.** All experimental protocols were approved by the Boston University Institutional Animal Care and Use Committee. GAD2-Cre/tdTomato mice (The Jackson Laboratory, ME, USA) were used for visualization of inhibitory interneurons. Brain slices were prepared from mice aged post-natal day 60 or greater (both genders). After anesthetization with isoflurane and decapitation, brains were removed and immersed in 0 °C solution of standard artificial cerebral spinal fluid (ACSF). For recordings, slices were moved to the stage of a two-photon imaging system. All recordings were conducted between 33 and 36 °C. Standard patch-clamp solutions and electrodes with resistances between 3 and 4 MΩ were used. The electrode pipette was visualized using the cyan-green fluorescent dye Alexa Fluor 488 hydrazide (Thermo Fisher Scientific Inc., MA, USA), which was added to the intracellular electrode solution (0.3% weight/volume). Imaging was performed using a two-photon imaging system (Thorlabs, Inc., NJ, USA) with a mode-locked Ti:Sapphire laser (Chameleon Ultra II; Coherent, CA, USA) set to wavelengths between 920 nm and 950 nm, which was used to excite both the Alexa Fluor 488 and tdTomato using a 20×, NA 1.0 objective lens (Olympus, MA, USA). Laser scanning was performed using resonant scanners and fluorescence was detected using two photomultiplier tubes (Hamamatsu, JP) equipped with red and green filters to separate emission from Alexa Fluor 488 and tdTomato. All other procedures were following our past work[49].

**Data analysis** Calcium images were analyzed using ImageJ. Calcium traces, electrophysiological traces were analyzed and plotted using Origin and MATLAB. All statistical analysis was done using Origin. Data shown are mean ± SD. *P* values were calculated based on two sample t-test and defined as: n.s.: non-significant, $p > 0.5$; *: $p < 0.5$; **: $p < 0.01$; ***: $p < 0.001$.

**Acknowledgements:** This work is supported by Brain Initiative R01 NS109794 to J-X.C. and C.Y. and National Institute of Health, United States, R01 NS052281 to J.A.W..

**Author contributions** C.Y., J-X.C., L.S., Y.J., and L.L. conceived the idea using taper fibers for high precision optoacoustic neurostimulation. L.S. and G.C. designed and fabricated the TFOE. L.S. and Y.J. performed the in vitro experiments. F.F., Y.J. and L.S. performed the patch clamp experiment. H-Y.M. provided neuron cultures. L.S., Y.J., J-X. C. and C.Y. wrote the manuscript. F.F., J. W. and H-Y.M. revised the manuscript. J.W. and H-Y.M. provided guidance for the project.

**Data Availability** The main data supporting the results in this study are available within the paper and its Supplementary Information. The raw data are available for research purposes from the corresponding author on reasonable request.